\documentclass[lettersize,journal]{IEEEtran}
\usepackage{amssymb}
\usepackage{caption}
\usepackage{graphicx}
\usepackage{xcolor}
\usepackage{subfig}
\usepackage{algorithm}
\usepackage{algpseudocode}
\usepackage{setspace}
\usepackage{float}
\usepackage{widetext}
\usepackage{flushend}
\usepackage{cuted}
\usepackage{cite}
\usepackage{comment}
\usepackage{amsfonts}
\usepackage{float}
\usepackage{amsmath}
\usepackage{subfig} 
\algnewcommand{\Initialize}[1]{%
  \State \textbf{Initialize:}
}
\algnewcommand{\Output}[1]{%
  \State \textbf{Output:}
}

\usepackage{amsmath}

\usepackage{mathtools}

\ifCLASSINFOpdf

\else

\fi

\begin{document}
%
\renewcommand{\thesubfigure}{} 
\title{BCH Coding Assisted Imaging}

\author{ Hao Jiang, Shuang Liu, Chentao Yue, Zihuai Lin

\thanks{Hao Jiang was with the School of Electrical and Computer Engineering, University of Sydney, Sydney, NSW 2006, Australia (e-mail: hjia2353@gmail.com).
Shuang Liu is with the Shanghai Institute of Optics and Fine Mechanics, Chinese Academy of Sciences (e-mail: liushuang0820@siom.ac.cn), contributed equally to this manuscript.
Chentao Yue and Zihuai Lin are with the School of Electrical and Computer Engineering, University of Sydney, Sydney, NSW 2006, Australia (e-mail: \{chentao.yue, zihuai.lin\}@sydney.edu.au).
}

}


\maketitle

\begin{abstract}
In modern correlation imaging systems, also known as ghost imaging (GI), particularly under low-light or noisy conditions, preserving high image fidelity presents a significant challenge. This paper introduces an innovative approach by integrating Bose-Chaudhuri-Hocquenghem (BCH) error control coding (ECC) into CGI systems to assist imaging. By encoding target image with BCH codes and using order-statistic decoding (OSD) for error correction during reconstruction, this approach significantly improves image quality across various signal-to-noise ratio (SNR) conditions. 
Simulation and experiment results validate that BCH coding assisted imaging achieves significantly enhanced robustness against additive white Gaussian noise (AWGN) and improved image reconstruction quality. In addition, the imaging performance of different BCH codes varies, with each code exhibiting distinct advantages based on factors such as code length and coding efficiency.

\end{abstract}


%
\IEEEpeerreviewmaketitle

\section{Introduction}
\label{intro}
\IEEEPARstart {G}{host} imaging reconstructs an unknown target by correlating a sequence of speckle patterns with the corresponding intensity measurements, where the bucket detector (BD) integrates the total incident power over its aperture \cite{Shih:08,Bornman01,Moreau01}. In many practical implementations, the speckle field is produced using pseudo-thermal illumination (e.g., a laser beam scattered by a rotating ground glass), which is well modeled as a Gaussian-state source and thus admits a statistical-correlation formulation \cite{Shapiro1,Erkmen:09,Pittman1995OpticalIB,Erkmen:10}. Moreover, it is shown that the performance of reconstructed images are sharply affected by the BD noise \cite{Shi:17}, such as thermal (Johnson–Nyquist) noise, are commonly modeled as signal-independent additive Gaussian noise \cite{PhysRev.32.97,PhysRev.32.110}.
\cite{Shapiro1,Erkmen:09,Pittman1995OpticalIB,Erkmen:10}.
Moreover, the reconstruction quality is significantly affected by the BD noise \cite{Shi:17}. In practice, BD noise is often dominated by electronic readout/amplifier noise and thermal (Johnson--Nyquist) noise; it is commonly modeled as signal-independent additive Gaussian noise \cite{PhysRev.32.97,PhysRev.32.110}.
Nevertheless, GI rarely include an explicit error-control mechanism to suppress or correct Gaussian-induced measurement perturbations\cite{wang2018error}. As a result, image fidelity can degrade significantly under strong Gaussian noise or transient disturbances. 
 In digital communication, by contrast, Gaussian channel impairments (often modeled as AWGN) are systematically addressed using ECC. Specifically, Shannon’s channel coding theorem establishes that arbitrarily reliable transmission is achievable whenever the coding rate is below the capacity of the Gaussian channel, and practical coded-modulation systems exploit this principle to realize substantial error-rate reductions in AWGN environments\cite{6773024,1697831,4282117}. Moreover, recent studies have shown that the above channel-coding-theoretic theorem is also applicable to GI \cite{Zhou:24}. Motivated by this connection, we introduce a channel-coding-based error-correction mechanism into the GI acquisition and reconstruction. In particular, we adopt BCH codes to enhance imaging performance. The features of BCH-based imaging, and the rationale for selecting BCH over other error-correcting codes, are discussed in Section II (B).

\textbf{Novelty and Contribution}: This paper incorporates channel coding into GI by encoding the target image in a BCH-based acquisition framework and reconstructing it via OSD. The proposed approach substantially improves robustness to AWGN and enhances reconstruction quality. The main contributions are summarized as follows.
\begin{enumerate}
    \item \textbf{Introduction of BCH Codes and OSD into GI}:  Building on the correlation-based GI framework, we construct deterministic illumination patterns from a BCH generator matrix to embed an explicit ECC structure into the imaging process. Rather than relying on conventional algebraic BCH decoders, we adopt OSD, which leverages soft reliability information in the measurements to estimate the transmitted codeword and, in turn, recover the target with improved robustness to Gaussian perturbations. 
    \item \textbf{Simulation and Experiment validation}: In a communication system, the distribution of the received signals is typically predictable. However, in an imaging system, the distribution of the received signals at the detector is more complex and diverse as a result of the noise. Based on the ordered illumination patterns, the range of received signals is estimated to simplify the theoretical analysis of OSD. Compared with the second-order correlation GI, the BCH-based imaging scheme achieves a clear performance improvement..
\end{enumerate}

\textbf{Outline of the Paper}: In Section II, we review the fundamentals of  GI and BCH-based imaging. Sections III and IV introduce the encoding and decoding processes of the BCH coding assisted imaging system. Section V provides a detailed theoretical analysis of the distribution of the received signal and reliability metrics, as well as the OSD performance in the imaging system. In Section VI , we present the simulation and experiment results. Finally, Section VII concludes the paper by summarizing the key findings and discussing future research directions.
\section{BCH-based Imaging system structure}
\subsection{Correlation Imaging System}
The imaging techniques are employed to to recover informative representations of a target, including its profile, location, and structural features \cite{abbas2024target,padgett2017ghost,strekalov2013ghost}. \textbf{Fig.~1(a)} illustrates a typical second-order correlation imaging (ghost imaging) configuration \cite{wang2018error,shapiro2012ghost}. A pseudo-thermal source is spatially modulated by a digital micro-mirror device (DMD) and then split into a test arm and a reference arm. \cite{Kassem:24,Pepe2020DistanceSO,Gong111,PhysRevA.87.023820,padgett2017ghost,Shapiro1,Erkmen:10,Bornman01}, in the test arm, the modulated field illuminates the target and is subsequently collected by a bucket detector (BD), which integrates the total transmitted/reflected intensity without spatial resolution. In the reference arm, the field is recorded by a charge-coupled device (CCD) to provide the spatially resolved speckle patterns. The target image is reconstructed by evaluating the second-order intensity correlation between the reference patterns and the corresponding bucket measurements. Notably, this correlation principle applies to both entangled-photon and pseudo-thermal illumination; in the entangled-photon case, the image can be inferred from correlation measurements even though the photons registered in the reference arm do not directly interact with the object\cite{Kassem:24,Pepe2020DistanceSO,Gong111}.

\subsection{BCH-based Imaging System} 
Prior work has demonstrated the feasibility of incorporating BCH coding into computational GI \cite{BAI2022108109}. Motivated by this, we adopt BCH coding to improve robustness against noise in BD, and further develop an OSD-based reconstruction/analysis framework. Importantly, although several error-correcting codes can introduce redundancy, BCH codes offer a set of properties that are well matched to the noise characteristics, decoding requirements, and practical constraints of GI systems. Accordingly, we adopt BCH coding in this work for three practical reasons. First, the dominant impairment in the considered GI measurements can be modeled as AWGN on intensity samples; after hard remapping into Galois Field
(\(\mathbf{GF^{(2)}}\)), the induced distortion is well approximated by random bit flips, for which BCH codes provide explicit multi-error-correction capability at short-to-moderate block lengths\cite{Berlekamp1965OnDB,1053833}. Second, BCH codes admit algebraic structure and short block lengths, which are essential in imaging systems where each codeword corresponds to a finite number of illumination patterns \cite{1057586,5075875}. In contrast, capacity-approaching codes such as LDPC or polar codes typically require long block lengths and iterative decoding \cite{4282117}, which are incompatible with the limited number of measurements and the strict latency constraints in practical GI systems. Third, BCH codes are particularly compatible with OSD, which exploits soft reliability information derived from intensity measurements. OSD provides near-maximum-likelihood performance for short linear block codes with manageable complexity, a property that is not readily available for Reed–Solomon codes (non-binary, higher complexity) or polar codes (successive cancellation structure, less effective for short lengths under non-standard soft metrics) \cite{fossorier1997soft,yue2022ordered,Van11}.

Fundamentally, imaging can be viewed as an information transfer process, closely tied to the principles of digital communication systems. 
Because of this, the imaging system can be simplified and modeled as a communication system, depicted in \textbf{Fig. 1(b)}\cite{Zhou:24,wang2018error}. As shown in \textbf{Fig. 1(c)}, both the communication system and the BCH-based imaging system share similar structures. This analogy underscores the shared principles of data transmission and processing between the two systems.
\begin{figure}
 \centering
\subfloat {
    \begin{minipage}{\columnwidth}
    \includegraphics[width=1\linewidth]{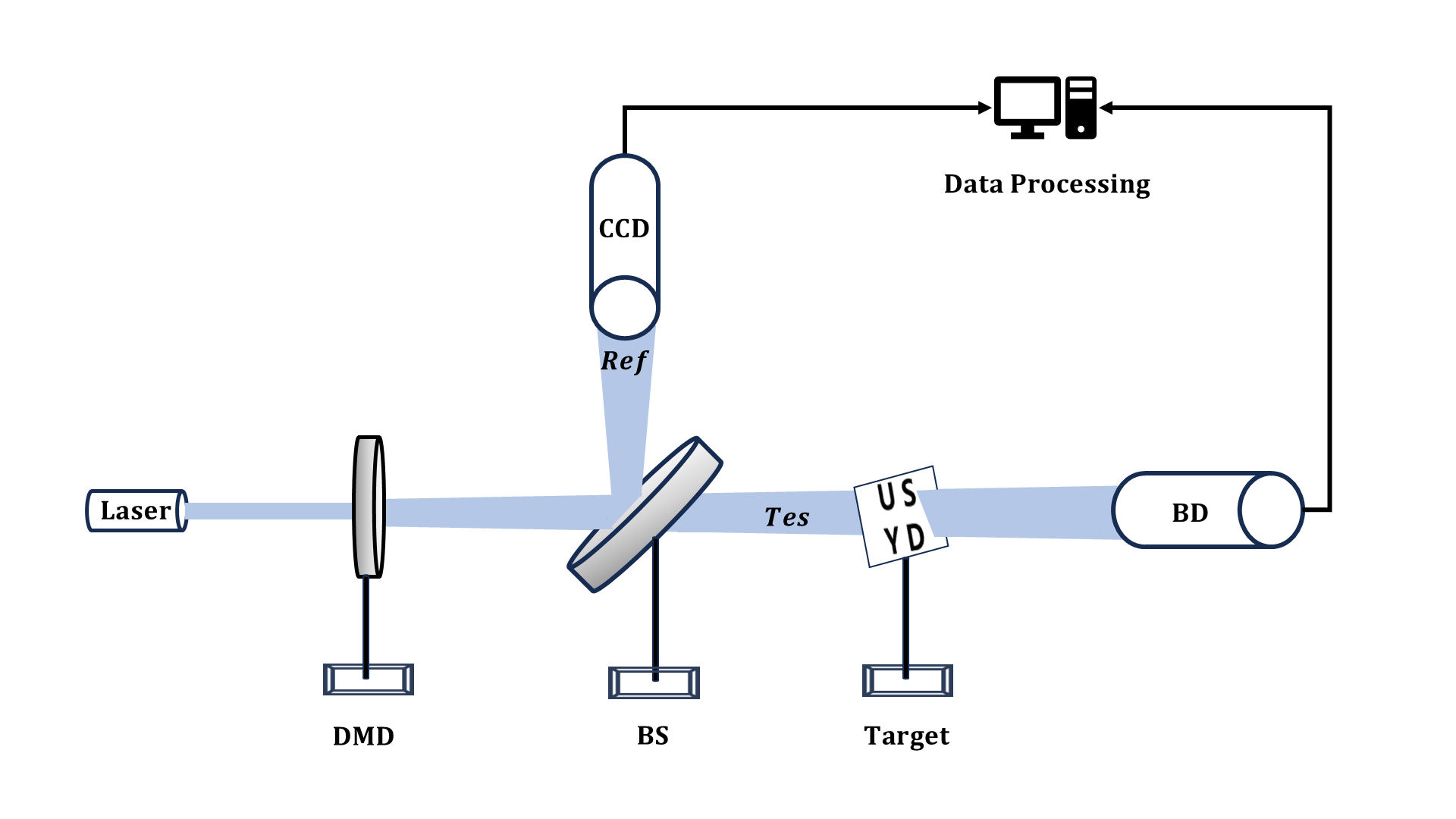}
    \textnormal{(a)}
    \centering
    \end{minipage}
} \\
\subfloat {
    \begin{minipage}{\columnwidth}
    \includegraphics[width=1\linewidth]{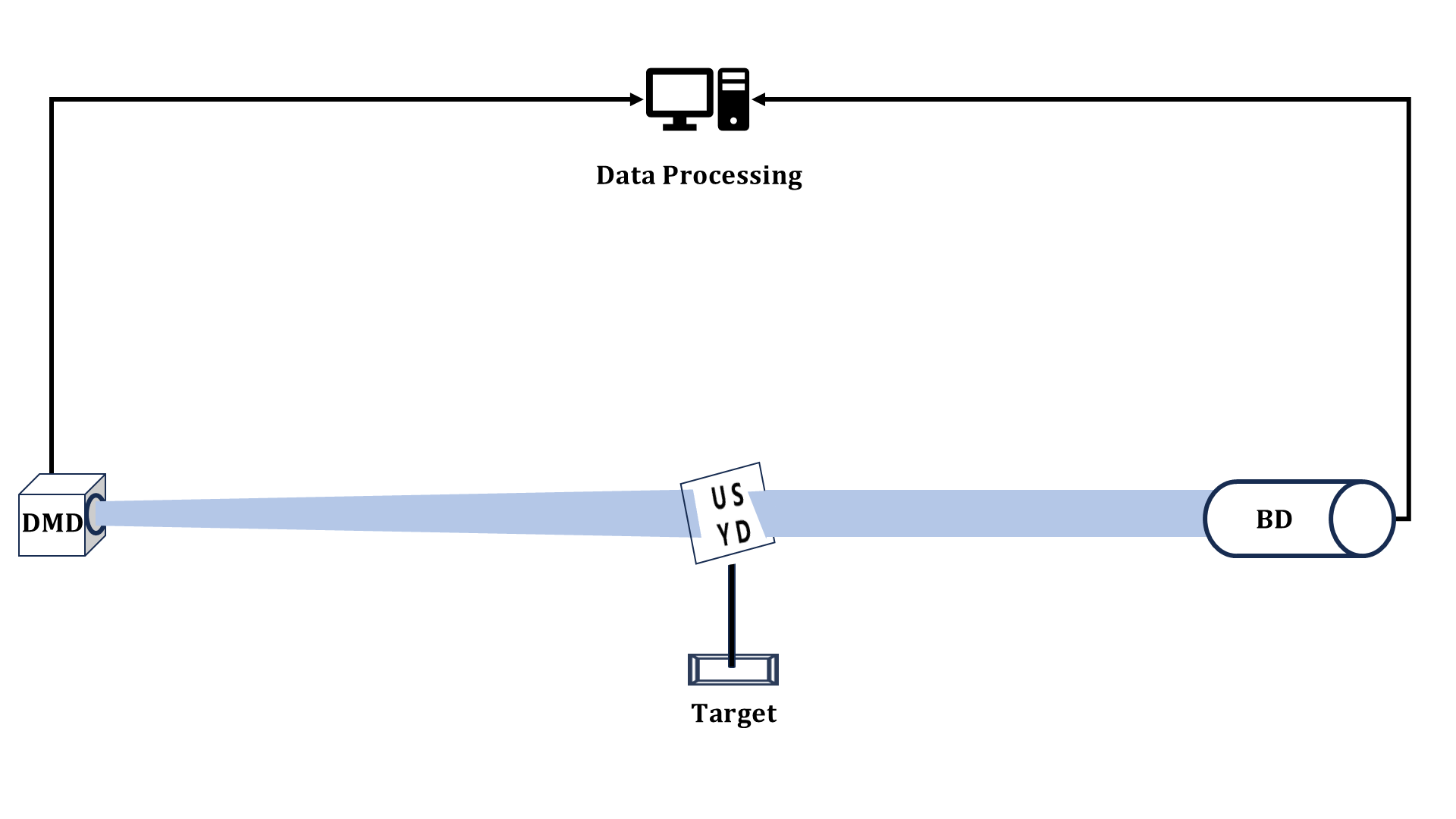}
    \textnormal{(b)}
    \centering
    \end{minipage}
} \\
\subfloat {
    \begin{minipage}{\columnwidth}
    \includegraphics[width=1\linewidth]{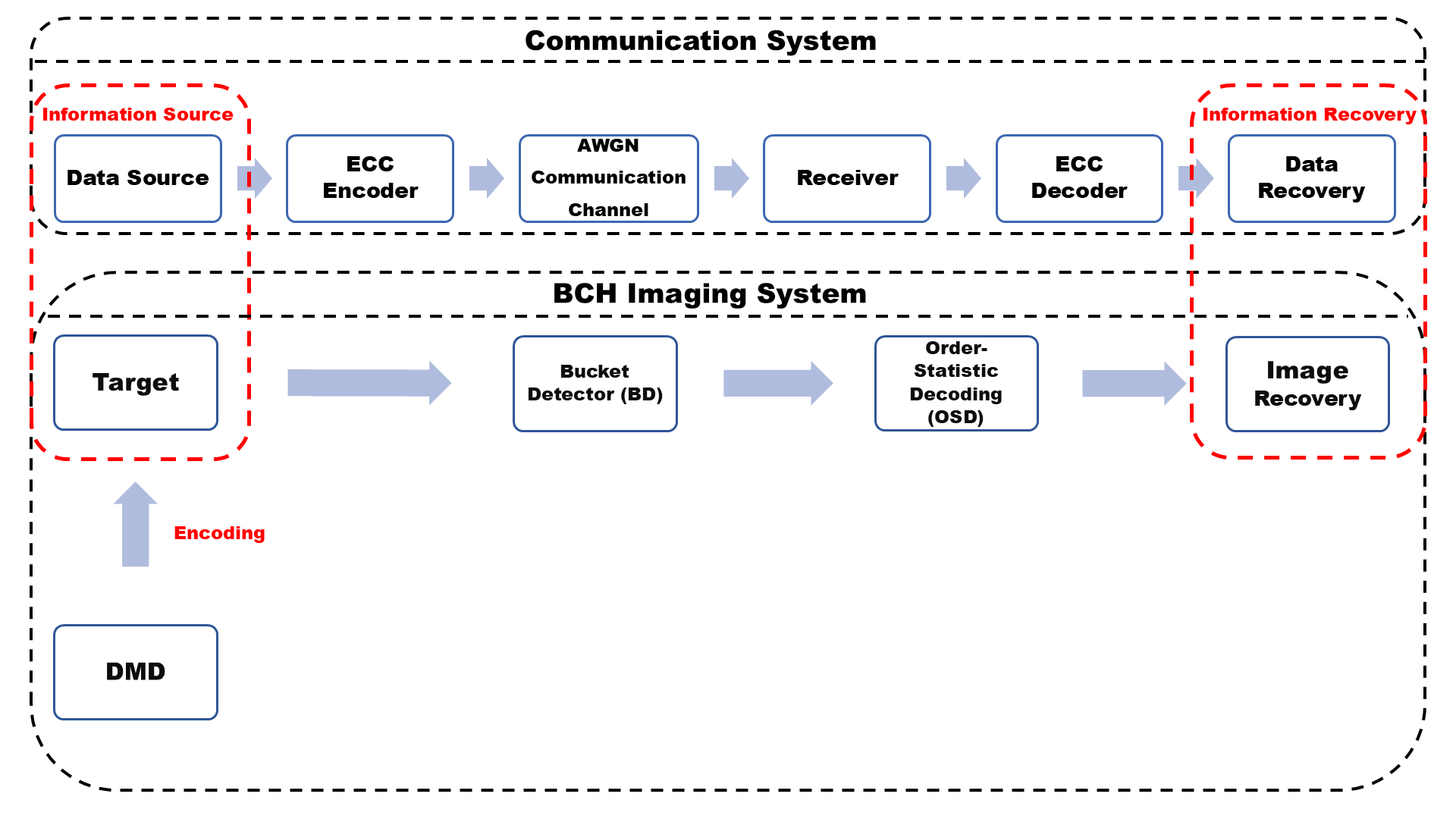}
    \textnormal{(c)}
    \centering
    \end{minipage} 
 }   
    \caption{The BCH-based imaging system is an evolution of the correlation imaging system and shares similarities with a digital communication system. (a). Second-order correlation GI system. (b). BCH-based imaging system. (c). A block diagram of a digital communication system and a BCH-based imaging system. Both systems include an information source, corresponding to the data source in a communication link and the target in an imaging system. In each case, information is conveyed through a channel — an engineered communication channel in the former and physical optical propagation in the latter. The transmitted information is then captured by a receiver (communication) or a BD (imaging) and subsequently processed to recover the original data stream or reconstruct the target image.}
\end{figure}

\section{Encoding and Decoding Process of  BCH-based Imaging}
In the context of our specific design, we simplify complex scenario by exclusively considering AWGN, explicitly neglecting two-dimensional ISI effects,turbulence and optical phase distortion. The BD is placed close to the target, and the signals intensity is predominantly influenced by AWGN alone, with propagation-induced signals attenuation, distortion, and other optical effects considered negligible. The encoding and decoding process of the proposed BCH-based imaging scheme is described as follows.

\subsection{Encoding Scheme}
As shown in \textbf{Fig. 2}, we adopt the \textbf{BCH (31, 16)} code for encoding. The corresponding generator matrix
$\mathbf{G}\in\{0,1\}^{K\times N}$ has size $16\times31$, where $N=31$ and $K=16$.
A $4\times4$ target image contains $16$ pixels and is represented in binary form, where pixel values
$0$ and $1$ denote black and white, respectively. The encoding procedure is summarized as follows.

First, we consider the systematic generator matrix $\mathbf{G}$ of the \textbf{BCH (31, 16)} code, denoted by
$\mathbf{G}=[\mathbf{g}_1,\mathbf{g}_2,\ldots,\mathbf{g}_N]$.
We take one column of $\mathbf{G}$, e.g.,
$\mathbf{g}_1=[1100100001110000]^T\in\{0,1\}^{16\times 1}$,
and reshape it into a $4\times 4$ matrix as:
\begin{equation*}
\hat{\mathbf{g}}_1 = \text{reshape}(\mathbf{g}_1, 4, 4) = 
\begin{bmatrix}
1 & 1 & 0 & 0 \\
1 & 0 & 0 & 0 \\
0 & 1 & 1 & 1 \\
0 & 0 & 0 & 0
\end{bmatrix}.
\end{equation*}
This reshaping enforces a one-to-one correspondence between the speckle-field structure and the target pixel layout, thereby facilitating efficient encoding and subsequent imaging.

Second, we employ an all-ones $4\times4$ binary target $\mathrm{T}$ (i.e., a white object) for modulation. 
For the $n$-th illumination, the interaction between the speckle pattern $\hat{\mathbf{g}}_n$ and the target 
produces a single bucket measurement, which is obtained by summing all element-wise products. For the all-ones target $\mathrm{T}$, the first received value \(i_t^{(1)}\) is:
\begin{equation}
\begin{aligned}
i_t^{(1)}
&=\sum_{u=1}^{4}\sum_{v=1}^{4}\left[\hat{\mathbf{g}}_1\right]_{u,v}\,T_{u,v} \\
&=\sum_{u=1}^{4}\sum_{v=1}^{4}\left[\hat{\mathbf{g}}_1\right]_{u,v}
=5,
\end{aligned}
\label{eq:it1_value}
\end{equation}

Repeating the above operation for all $N$ illuminations yields the measurement vector:
\begin{equation}
\mathbf{I}_\mathbf{t}=\left[i_t^{(1)},\, i_t^{(2)},\, \ldots,\, i_t^{(N)}\right].
\label{eq:It_def}
\end{equation}
The sequence $\mathbf{I}_\mathbf{t}$ collects the bucket intensities corresponding to all speckle patterns, 
thereby encoding the target information into a set of scalar measurements for subsequent reconstruction.

\begin{figure}
    \centering
    \includegraphics[width=1\linewidth]{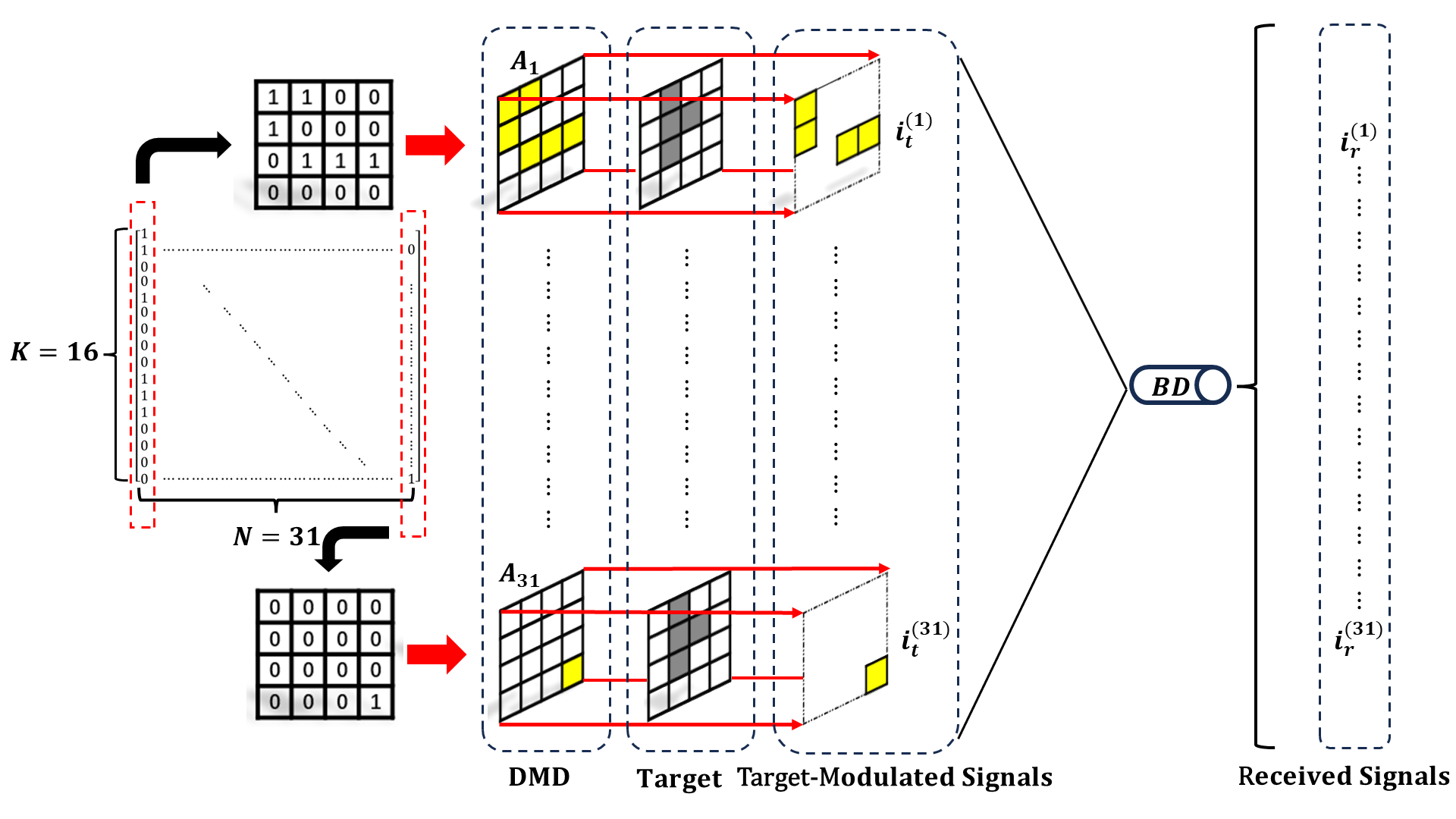}
    \caption{An example of imaging with \textbf{BCH (31, 16)} involves the generation of a speckle field using a DMD, which is determined by the generator matrix of the BCH code. The speckle field illuminates the target, and the total intensity of the target-modulated signals is collected by BD. This process integrates the principles of BCH coding with imaging. }
    \label{fig_2}
\end{figure}

\subsection{Principle of Image Reconstruction}
Let the generator matrix of a BCH$(N,K)$ code be written in systematic form as
$\mathbf{G}=[\mathbf{Q}\,|\,\mathbf{I}_K]$,
where $\mathbf{I}_K$ is the $K\times K$ identity matrix and $\mathbf{Q}$ is a $K\times(N-K)$ submatrix.
Pixel-wise scanning is employed to ensure that the target information is fully acquired. As discussed in Section~III,
the identity part $\mathbf{I}_K$ provides $K$ elementary illumination patterns that enable direct pixel scanning,
whereas $\mathbf{Q}$ introduces redundancy into the codeword, analogous to its role in digital communication systems.

Nevertheless, point-wise scanning can be highly sensitive to noise fluctuations,
since the measurement associated with a single pixel (or a small set of pixels) may suffer from low energy and thus a low SNR.
To alleviate this vulnerability, we apply invertible row/column transformations to the generator matrix \(\mathbf{G}\) to redistribute
and effectively average the illumination energy across patterns, thereby improving the robustness of the measurements. 
Because these transformations are reversible, the corresponding inverse transformation is applied in decoding,
so that the original codeword structure is preserved and the target image can be correctly reconstructed.

\subsection{Modulo-2 operations Substitution and mapping rules}
Unlike digital communication systems, the imaging system lacks structures to directly perform \textbf{XoR} operations. However, it is still possible to map the received signals in the \(\mathbf{GF^{(2)}}\) using log-likelihood ratios (\textbf{\textit{L}})\cite{wu2002likelihood,choi2011geometry}. In our scenario, the dominant factor degrading the reconstruction quality is the AWGN introduced at the BD. The relationship between the received and modulated signals is given as \textbf{\(\textbf{I}_\textbf{r}=\textbf{I}_\textbf{t}+\boldsymbol{\Omega}\)}, where \(\boldsymbol{\Omega}=[\omega^{(1)},\omega^{(2)}...\omega^{(N)}]\) represents the AWGN sequence. The log-likelihood ratios (\textbf{\textbf{\textit{L}}}) for the received signals \(\textbf{I}_\textbf{r}\) are calculated using the formula: 
\begin{equation} 
\textbf{\textit{L}} = \ln \left[ \frac{P_{{I_{r\text{max} }|I_r}}}{P_{{I_{r\text{min} }|I_r}}} \right]
\end{equation} 
Here, \(\textbf{\textit{L}}=[\ell^{(1)},\ell^{(2)}...\ell^{(N)}]\), where each component 
\(\ell^{(n)}\) is the log-likelihood ratio for a specific received signal.
\begin{itemize}
    \item  \(\textbf{I}_{\textbf{rmax}}=[i_{rmax}^{(1)},i_{rmax}^{(2)}...i_{rmax}^{(N)}]\) is the upper sideband sequence.

    \item \(\textbf{I}_{\textbf{rmin}}=[i_{rmin}^{(1)},i_{rmin}^{(2)}...i_{rmin}^{(N)}]\) is the lower sideband sequence.
\end{itemize}
This method ensures that the signals can be processed effectively in the absence of direct \textbf{XoR} operations, enabling robust decoding despite the impact of noise.
The received signal \(i_r^{(n)} \in \textbf{I}_\textbf{r}\) is mapped in the \(\mathbf{GF^{(2)}}\) using Algorithm 1. 

\begin{algorithm}[H]
\caption{Received Signal Mapping Into $\mathbf{GF^{(2)}}$}
\label{alg:alg1}
\begin{algorithmic}[1]
\Require Received sequence $\mathbf{I}_r = [I_r^{(1)},\dots,I_r^{(N)}]$, noise std. $\sigma$
\Ensure Hard decision sequence  $\mathbf{I_r^{\textbf{GF(2)}}} \in \{0,1\}^N$, Log-likelihood ratio (LLR) $\textbf{\textit{L}}\in\mathbb{R}^N$, candidate matrix $\mathbf{R}\in\mathbb{Z}^{2\times N}$, chosen sequence $\mathbf{c}\in\mathbb{Z}^N$

\State $N \gets \text{numel}(\mathbf{I}_r)$
\State $\mathbf{R} \gets \textsc{Range}(\mathbf{I}_r)$ \Comment{$\mathbf{R}(1,n)=i_{rmax}^{(n)}$, $\mathbf{R}(2,n)=i_{rmin}^{(n)}$}
\State $\textbf{\textit{L}} \gets \textsc{Get\_LLR}(\mathbf{I}_r,\mathbf{R},\sigma)$
\Comment{Refer to formula (3)}
\For{$n=1$ to $N$}
    \If{$\ell^{(n)} > 0$}
        \State $c^{(n)} \gets i_{rmax}^{(n)} \; (= \mathbf{R}(1,n))$
    \Else
        \State $c^{(n)} \gets i_{rmin}^{(n)} \; (= \mathbf{R}(2,n))$
    \EndIf
    \State $\mathbf{I_r^{\textbf{GF(2)}}(n)} \gets c^{(n)} \bmod 2$
    \Comment{even $\rightarrow 0$, odd $\rightarrow 1$}
\EndFor

\State \Return $\mathbf{I_r^{\textbf{GF(2)}}}$, $\textbf{\textit{L}}$
\end{algorithmic}
\end{algorithm}
After this mapping, the resulting binary sequence \(\mathbf{I_r^{\textbf{GF(2)}}}\) is the hard decision sequence, expressed as: 
\begin{equation}
 \mathbf{I_r^{\textbf{GF(2)}}}=\mathbf{[\textit{i}_\textit{r}^{\textit{GF(2)}}(1),\textit{i}_\textit{r}^{\textit{GF(2)}}(2)...\textit{i}_\textit{r}^{\textit{GF(2)}}(N)]}
\end{equation}

Both sequences, \(\textbf{I}_{\textbf{rmax}}\) and \(\textbf{I}_{\textbf{rmin}}\), are visually represented in \textbf{Fig. 3(a)}. 
\textbf{Fig. 3(b)} illustrates the proposed mapping rule. 
The received measurements are first quantized by thresholding at the midpoints between adjacent integer levels, i.e., each sample is rounded to its nearest integer. The resulting integers are then converted to \(\mathbf{GF^{(2)}}\) symbols according to the predefined mapping table.

It is worth noting that as the noise power (\(\sigma^2\)) increases, the intersections between different curves of the PDF become larger. This results in increased decision ambiguity, introducing more errors into the remapping process. When the received signals are near the intersections, it becomes more difficult to accurately map them to the intended values. 
As a result, once \(\textbf{I}_\textbf{r}^{\textbf{GF(2)}}\) is obtained, the OSD algorithm is used to mitigate errors and recover the target information. The OSD algorithm operates as follows:
\begin{enumerate}
    \item Analyzing Reliability: 
         Refer to \(\textbf{\textit{L}}\), assessing the reliability of each bit in the binary sequence \(\textbf{I}_\textbf{r}^{\textbf{GF(2)}}\).
    \item Soft Decision-Making: Using reliability metrics to identify and correct potential errors.
    \item Iterative Refinement: Ensure that the decoded information closely matches the original transmitted data.
\end{enumerate}
\begin{figure}
\centering
 \subfloat[(a) ]
 { \centering
 \includegraphics[width=1\linewidth]{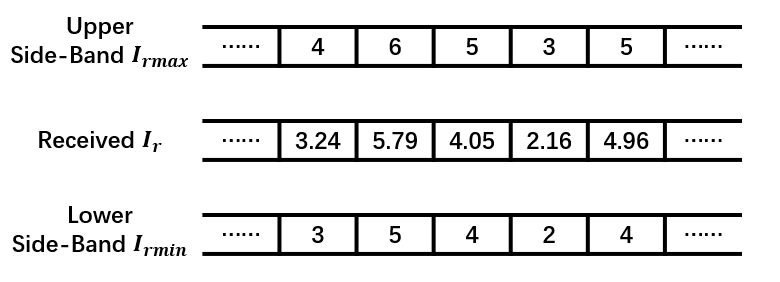}}\\
\subfloat[(b)] { 
\centering \includegraphics[width=1\linewidth]{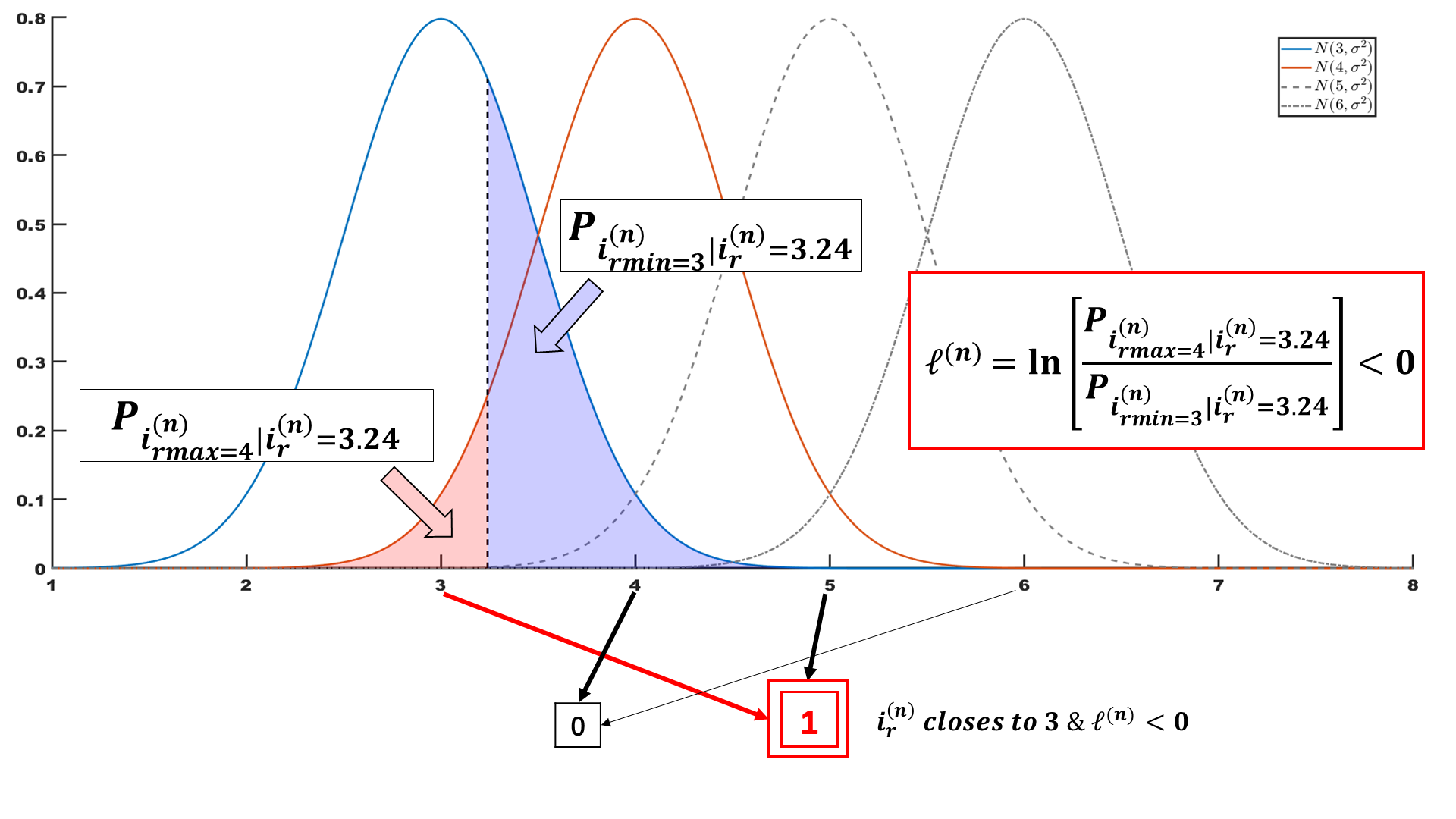}}\\
\caption{
The PDF of the received signals and its remapping rules. (a) Find two closest integers to \(\textit{i}_\textit{r}^{(n)}\). (b) The red and blue shaded regions correspond to the conditional probability \(P_{i_t^{(n)}|i_r^{(n)}}\), with larger areas indicating higher probabilities; accordingly, \(i_t^{(n}\) is restricted to the two most likely integer values, denoted by \(i_{rmax}^{(n)}\) and \(i_{rmin}^{(n)}\). For example, at \(\textit{i}_\textit{r}^{(n)}=3.24\), the log-likelihood ratio satisfies \(\ell^{(n)}<0\). As a result, \(\textit{i}_\textit{r}^{(n)}=3.24\) is mapped to the bit value 1.}
\label{fig_4}
\end{figure}

The OSD algorithm leverages soft-decision techniques, which make it highly effective in handling noise-induced errors, ensuring accurate recovery of the target information even in challenging noise environments\cite{xu2012osd,fossorier1997soft}.

\subsection{Order-Statistic Decoding Algorithm}
\subsubsection{\(\textit{I}_\textbf{\textit{r}} \) Ordered by Reliability} 
For the binary sequence \(\mathbf{I_r^{\textbf{GF(2)}}}\) and the reliability metrics \(|\textit{\textbf{L}}|= \mathbb{L}\) , the relationship can be expressed as:
\begin{equation}
\begin{aligned}    \mathbf{I_r^{\textbf{GF(2)}}}=&\mathbf{[\textit{i}_\textit{r}^{\textit{GF(2)}}(1),\textit{i}_\textit{r}^{\textit{GF(2)}}(2)...\textit{i}_\textit{r}^{\textit{GF(2)}}(n)]}\\
     &\Updownarrow\\
    |\textbf{\textit{L}}|=&[|l^{(1)}|,|l^{(2)}|...|l^{(n)}|]\\
     &\Updownarrow\\
     \mathbb{L}=&[\ell^{(1)},\ell^{(2)}...\ell^{(n)}]
     \end{aligned}
\end{equation}
Here, \(|\cdot|\) represents the absolute operator, and \(\Updownarrow\) indicates that the elements of \(\mathbf{\textbf{I}_\textbf{r}^{\textbf{GF(2)}}}\) are mapped to the corresponding elements in \(|\textbf{\textit{L}}|\) where \(|l^{(n)}|\) is denoted by \(\ell^{(n)}\). 

In the OSD decoding process, \(\mathbf{I_r^{\text{\textit{GF(2)}}}}\) is sorted in descending order of reliability \(\mathbb{L}\)\cite{xu2012osd,fossorier1997soft,yue2022ordered,fossorier1995soft}. This sorting operation can be represented as: 
\begin{equation} 
 \hat{\mathbb{V}}= \xi_1 [I_r^{\text{\textit{GF(2)}}}]\end{equation} 
 where \(\xi_1[\cdot]\) denotes a permutation operator. To ensure consistency, the reliability vector \(\mathbb{L}\) and the corresponding columns of the generator matrix \(G\) are reordered in the same way as \(\textbf{I}_\textbf{r}^{\textbf{GF(2)}}\). The sorted reliability vector and matrix are expressed as: 
 \begin{equation}    
\begin{aligned}
  \hat{\mathbb{L}} \ = \xi_1[\mathbb{L} ]\\ 
  \hat{\mathbb{G}} = \xi_1 [G ]
\end{aligned} 
\end{equation}
Here, \( \hat{\mathbb{L}}\) is the reliability vector after sorting, and \(\hat{\mathbb{G}}\) is the generator matrix reordered according to \(\xi_1[\cdot]\). These steps are critical to aligning the decoding process with the reliability-based sorting, ensuring optimal error correction performance. 
\subsubsection{Gaussian Elimination in OSD Decoding} 
Gaussian elimination is used to solve systems of linear equations through row reduction. To ensure that the first \textit{k} columns of \(\hat{\mathbb{G}}\) are linearly independent after the first permutation (\(\xi_1\)), an additional permutation \(\xi_2[\cdot]\) is applied\cite{yue2022ordered,misra1996order,mcginn2002parallel}. This results in updated sequences and matrices as follows: 
\begin{equation}
\begin{aligned}
\mathbb{V}^*&=\xi_2[\hat{\mathbb{V}}]=\xi_2[\xi_1 [I_r^{\text{\textit{GF(2)}}}]]\\
\mathbb{L}^* &= \xi_2[\hat{\mathbb{L}}]=\xi_2[\xi_1[\mathbb{L} ]]\\
\mathbb{G}^* &= \xi_2[\hat{\mathbb{G}}]=\xi_2[\xi_1 [G ]]
\end{aligned}
\end{equation}

After applying both permutations (\(\xi_1\) and \(\xi_2\)), the binary sequence \(\mathbb{V}^*\) and its associated reliability vector \(\mathbb{L}^*\) are reordered such that the first \textit{k} positions in \(\mathbb{V}^*\) correspond to the most reliable bits\cite{yue2022ordered}. The matrix \(\mathbb{G}^*\) is similarly reordered, ensuring that its first \textit{k} columns are linearly independent. The first \textit{k} positions of  \(\mathbb{V}^*\), denoted as \(\mathbb{V}^*_k=\mathbb{V}^*[1:k]\).
This arrangement prioritizes the most reliable bits, minimizing errors and improving the accuracy of subsequent decoding steps. The reordering process ensures alignment between the binary sequence, reliability metrics, and the generator matrix for effective error correction. 
\subsubsection{Testing Error Patterns (TEPs) }
The process of Testing Error Patterns (TEPs) involves evaluating possible error patterns \( \mathbb{P_e}\) applied to \(\mathbb{V}^*_k\) to generate valid codeword through re-encoding\cite{yagi2005heuristic,dhakal2016error}. This is expressed as: 
\begin{equation}
 V^* = [ \mathbb{V}^*_k\oplus \mathbb{P_e}] \mathbb{G}^* 
 \end{equation}
where \( V^*\) represents the estimated codewords after encoding. Here, \(\oplus\) denotes the \textbf{XoR} operation. The algorithm systematically tests \(2^k\) possible error patterns, projecting each \( \mathbb{P_e}\) onto the hard-decision sequence \(\mathbb{V}^*_k\) \cite{yue2022ordered,li2023preconfigured}. 

The error patterns are tested in increasing order of the Hamming weight (\(d_H\)). Testing all possible error patterns with \(d_H<M\) is referred to as \(M\)-reprocessing. While testing all \(2^k\) patterns corresponds to Maximum Likelihood (ML) decoding, this is computationally infeasible for codes. To address this, the algorithm seeks the codewords \( V^*\) at a minimum Euclidean distance from \(\hat{\mathbb{V}}\)\cite{fossorier1997soft,fossorier1995soft,yue2022ordered}. The resulting codeword, \(V^*_{out}\), is selected as the most likely received codeword. 

In OSD, the goal is to reduce computational complexity while maintaining robust error correction performance. By minimizing unnecessary error pattern generation, OSD achieves efficient decoding. Simulations demonstrate that for block codes of length \(n\leq64\), \(2\)-reprocessing achieves performance comparable to the ML decoding algorithm\cite{fossorier1997soft,fossorier1995soft,yue2022ordered}. The final estimated codeword \(\vec V_{out}\) is obtained by reversing the permutations applied during the decoding process:
\begin{equation}
\vec V_{out} = \xi_1^{-1} \left[ \xi_2^{-1} \left[ V^*_{out} \right] \right]
\end{equation}
Where \( \xi_1^{-1} \)and \( \xi_2^{-1} \) are the inverse permutation applied during the reordering steps\cite{fossorier1997soft,fossorier1995soft,yue2022ordered,yagi2005heuristic,dhakal2016error}. This process ensures that the decoded codeword aligns with the original order, providing a high-quality estimate of the transmitted data while maintaining computational efficiency.

\section{Theoretical Analysis of  Imaging Performance}
For simplicity and without loss of generality, the target is assumed to be an blank image (fully transmissive), such that all light modulated by the DMD passes through the target. Each speckle element is normalized to unit intensity. We employ a \textbf{BCH (31, 16)} coding scheme for coded illumination, and the n-th measurement \(i_r^{(n)}\) is given by \(i_r^{(n)}=i_t^{(n)}+\omega^{(n)}\). We analysis the statistical distribution of \(\textbf{I}_\textbf{r}\) and \(\mathbb{L}\), and then theoretically derive the OSD performance of the proposed BCH-based imaging scheme.
\subsection{Distribution Analysis of Received Signals and Reliability Metrics}
 Under the blank-target assumption, the BD output equals the number of active speckle elements in the \(n\)-th illumination pattern. Consequently, the BD produces only three possible noiseless values, \(\{1,7,11\}\), corresponding to illumination patterns containing 1, 7, and 11 active speckles, respectively. The corresponding probabilities for \(i_t^{(n)} \in \{1,7,11\}\) are: \(P_{1} = \frac{16}{31}\), \(P_{7} = \frac{10}{31}\), \(P_{11} = \frac{5}{31}\).

 The PDF of \(\textbf{I}_\textbf{r}\) is a mixture of Gaussian distributions: 
 
\begin{equation}
    f_{\textbf{I}_\textbf{r}}(i_r^{(n)}) = \sum_{j \in \{1, 7, 11\}}[P_{j} \cdot \frac{1}{\sqrt{2\pi{\sigma}^2}} e^{-\frac{(i_r^{(n)} - {i_t^{(n)}})^2}{2\sigma^2}}]
\end{equation}
where:
\begin{itemize}
   \item \(\sigma^2=10^{\mathrm{-SNR_{dB}}/10}\).
   \item \(\mathrm{SNR_{dB}}\) denotes the BD signal-to-noise ratio.
\end{itemize}

To determine the distribution of \(\textbf{L}\), the calculation should rely on the closest distribution of \(\textbf{I}_{\textbf{rmax}}\) and \(\textbf{I}_{\textbf{rmin}}\) surrounding the received signals \(\textbf{I}_\textbf{r}\). For any \(i_r^{(n)}\), equation (3)  can be denoted as:
\begin{equation}
\begin{aligned}
   L(n) = &\ln \frac{\sum_{j \in \{1, 7, 11\}}[P_{j} \cdot \frac{1}{\sqrt{2\pi{\sigma}^2}} e^{-\frac{(i_r^{(n)} - {i_{rmax}^{(n)}})^2}{2\sigma^2}}]}{\sum_{j \in \{1, 7, 11\}}[P_{j} \cdot \frac{1}{\sqrt{2\pi{\sigma}^2}} e^{-\frac{(i_r^{(n)} - {i_{rmin}^{(n)}})^2}{2\sigma^2}}]}\\
   \approx&\ln { e^{-\frac{(i_r^{(n)}-i_{rmax}^{(n)})^2}{2\sigma^2}} } - \ln { e^{-\frac{(i_r^{(n)}-i_{rmin}^{(n)})^2}{2\sigma^2}} }\\
\approx&\frac{i_r^{(n)} -\Delta \alpha/2}{\sigma^2} 
\end{aligned}
\end{equation}
Where:
\begin{itemize}
    \item \(\Delta \alpha= i_{tmax}^{(n)}+i_{min}^{(n)}\).
    \item Defined as \( \beta= i_r^{(n)}-\Delta \alpha/2\), where \(\beta \in \{-0.5,0.5\}\).
\end{itemize}
When \(i_r^{(n)} \in (0,1)\), \(\Delta \alpha/2 = 0.5\); \(i_r^{(n)} \in (1,2)\), \(\Delta \alpha/2 = 1.5\). The PDF of log-likelihood ratios  \({L}_1\) is derived as:

\begin{equation}
    f_{L_1}(l^{(n)}) = \frac{1}{2} \cdot \frac{{\sigma}}{\sqrt{2\pi}} e^{-\frac{{\sigma}^2(i_r^{(n)} - \frac{1}{{2\sigma}^2})^2}{2}}+\frac{1}{2} \cdot \frac{{\sigma}}{\sqrt{2\pi}} e^{-\frac{{\sigma}^2(i_r^{(n)} + \frac{1}{{2\sigma}^2})^2}{2}}
\end{equation}
The PDF of log-likelihood ratios  \(\mathbb{L}_1\) is derived as:
\begin{equation}
\begin{aligned}
& f_{\mathbb{L}_1}(\ell^{(n)}) = \\
& \begin{cases} 0 & {\text{if}\ \ell^{(n)}>0}\\
{\frac{{\sigma}}{\sqrt{2\pi}} e^{-\frac{{\sigma}^2(i_r^{(n)} - \frac{1}{{2\sigma}^2})^2}{2}}+ \frac{{\sigma}}{\sqrt{2\pi}} e^{-\frac{{\sigma}^2(i_r^{(n)} + \frac{1}{{2\sigma}^2})^2}{2}}}, &{\text{if}\ \ell^{(n)}>0}
    \end{cases} 
\end{aligned} 
\end{equation}
Similarly, the PDFs of the log-likelihood ratios can be derived for other decision intervals, including \({L}_2\) for \(i_r^{(n)} \in (6,8)\); \({L}_3\) for \(i_r^{(n)} \in (10,12)\). Accordingly, the corresponding \(\mathbb{L}_2(\ell^{(n)})\) and \(\mathbb{L}_3(\ell^{(n)})\) can also be obtained.
Finally, the intergrade PDF of \(\mathbb{L}\) is derived as:
\begin{equation}
\begin{aligned}
 f_{\mathbb{L}}(\ell^{(n)}) &= P_1 \cdot f_{\mathbb{L}_1}(\ell^{(n)})+P_2 \cdot f_{\mathbb{L}_2}(\ell^{(n)})+P_3 \cdot f_{\mathbb{L}_3}(\ell^{(n)}) \\
 &=\frac{{\sigma}}{\sqrt{2\pi}} e^{-\frac{{\sigma}^2(i_r^{(n)} - \frac{1}{{2\sigma}^2})^2}{2}}+ \frac{{\sigma}}{\sqrt{2\pi}} e^{-\frac{{\sigma}^2(i_r^{(n)} + \frac{1}{{2\sigma}^2})^2}{2}}
\end{aligned} 
\end{equation}
Where, \(\ell^{(n)}>0\).

\subsection{Analysis of  OSD Performance in Imaging System}
The Q-function can be defined by \(Q(x)=\frac{ 1}{ \sqrt{2\pi }}\int_{x}^{\infty} e^{-\frac{u^2}{2}}du\) , the cumulative distribution function (CDF) of \(\mathbb{L}\) can be derived as\cite{yue2022ordered,mukherjee2020study,maiti2016estimators}:
\begin{equation} 
\begin{aligned}
&F_{ \mathbb{L}} (\ell^{(n)})= \\
&\begin{cases} 
0,  & \ell^{(n)} < 0\\
1-Q \left[{\sigma}(\ell^{(n)}+\frac{1}{2{\sigma}^2})\right]-Q \left[{\sigma}(\ell^{(n)}-\frac{1}{2{\sigma}^2})\right],&\ell^{(n)} > 0
\end{cases}
\end{aligned}
\end{equation}

The second permutation in Gaussian elimination is omitted, the PDF of the \(n-th\) order reliability \(\mathbb{L}^*\)can be derived in as \cite{papoulis2002probability}:

\begin{equation}
\begin{aligned}
f_{\mathbb{L}^*}({\ell^*}^{(n)})&= \frac {N!} {(n- 1)! (N - n)!} \\
&\cdot(1 - F_\mathbb{L}({\ell^*}^{(n)}))^{n-1}  F_\mathbb{L}({\ell^*}^{(n)})^{N-n} f_\mathbb{L} ({\ell^*}^{(n)})
\end{aligned}
\end{equation}
Using \(E(a,b)\) to denote the number of errors in the positions \(a\) to \(b\), the probability mass function (PMF) \(P_{E(a,b)}\), which gives the probability that a discrete random variable is exactly equal to some value\cite{stewart2011probability}
, is given by \cite{9427228}:

\begin{equation}
    P_{E(a,b)}=\int_{0}^{\infty} \binom {b}{j} p( x)^j (1-p( x))^{b-j} f_{|\mathbb{L}^*|_{b+1}}(x)dx  
\end{equation}
Where, \(a=1 <b < N \), \(0\leq j\leq b-a+1\). 

In our design, \(p(x) \) is derived as:
\begin{equation}
     p(x)=   \frac{Q [\hat{\sigma}_1(x+\frac{1}{2\sigma^2})]}{1+Q [\hat{\sigma}_1(x+\frac{1}{2\hat{\sigma}_1^2})]-Q [\hat{\sigma}_1(-x+\frac{1}{2\hat{\sigma}_1^2})]}
 \end{equation}
Equation (19) can describe the probability mass function (PMF) of the number of errors over the ordered hard-decision \(i^*_r(n)\). Equations (17)-(19) were investigated in a previous work \cite{yue2022ordered}[Eq. (15)-(17)]. \\
\begin{figure}
    \centering
\subfloat[(a) ]
    {\includegraphics[width=1\linewidth]{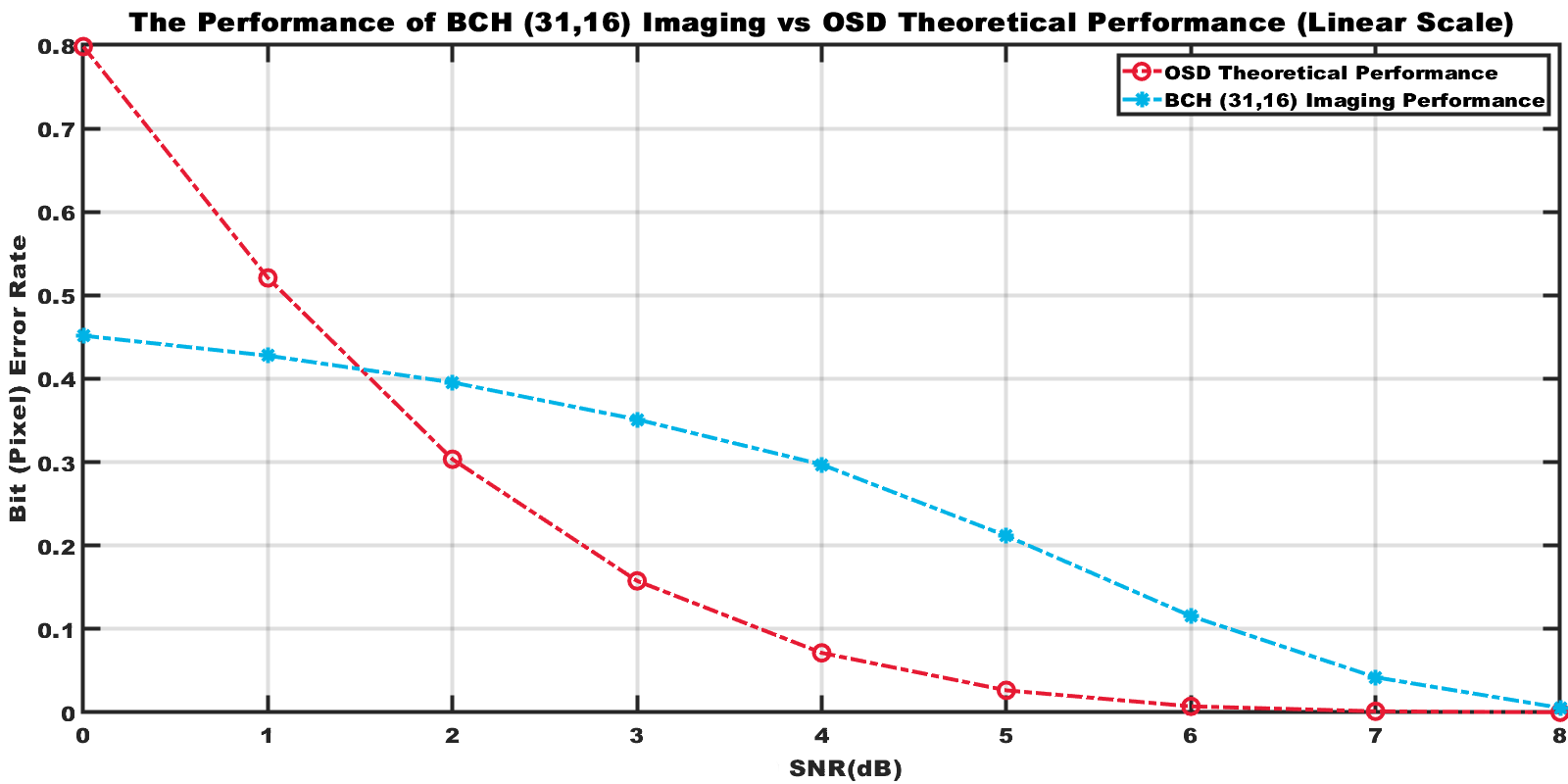}}\\
\subfloat[(b) ]
    {\includegraphics[width=1\linewidth]{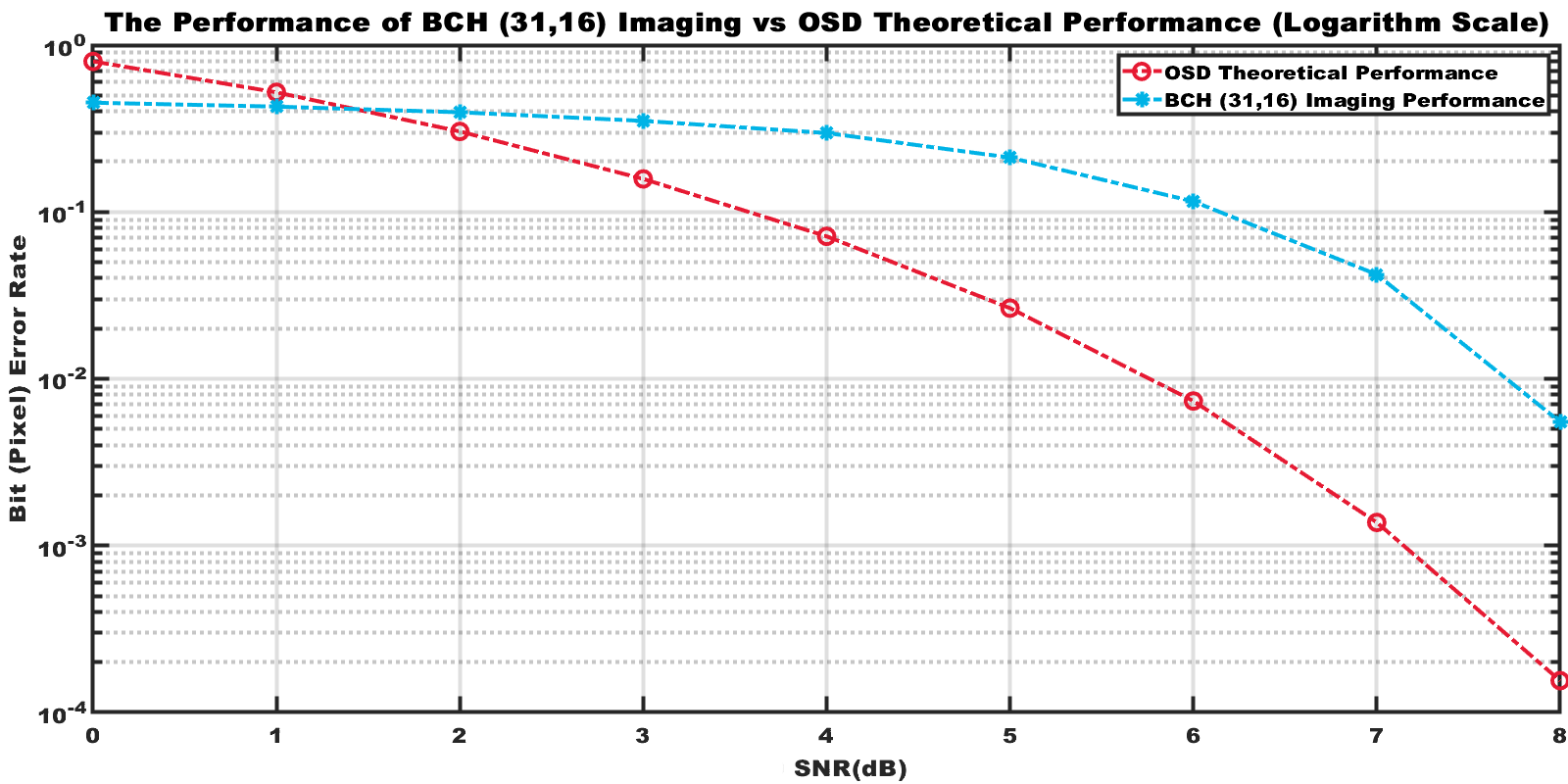}}\\
    \caption{The performance of \textbf{BCH (31, 16)} imaging versus the OSD theoretical prediction, shown in linear scale in (a) and logarithmic scale in (b). When the BD SNR exceeds \(9\,\mathrm{dB}\), the simulations indicate that \textbf{BCH (31, 16)} achieves error-free reconstruction of the target. } 
    \label{fig:placeholder}
\end{figure}

\textbf{Fig. 4} illustrates the simulation performance of \textbf{BCH (31, 16)} imaging and the corresponding theoretical bit-error-rate 
\begin{figure}
    \centering
\subfloat[(a) ]
    {\includegraphics[width=1\linewidth]{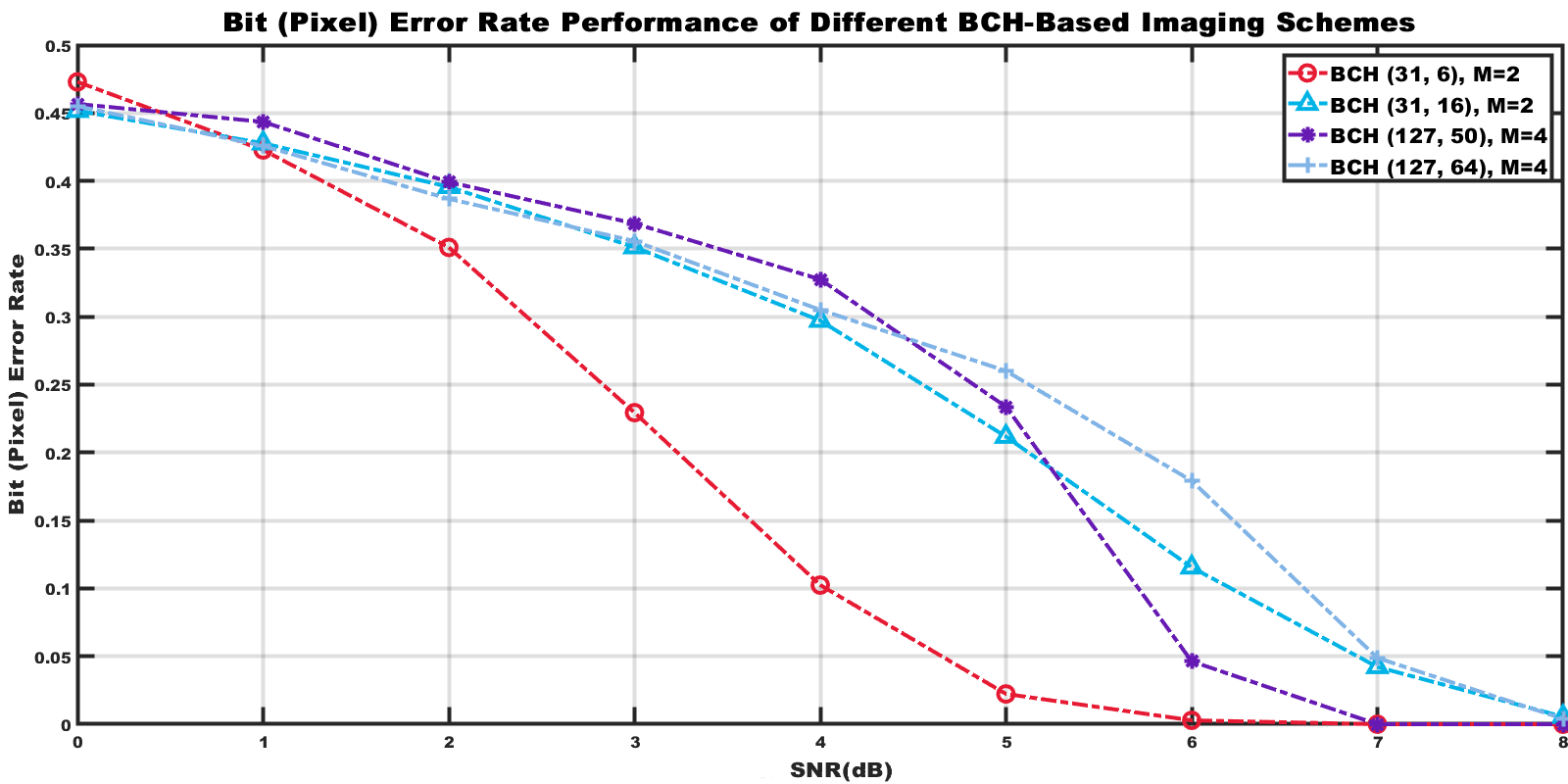}} \\
\subfloat[(b) ]
    {\includegraphics[width=1\linewidth]{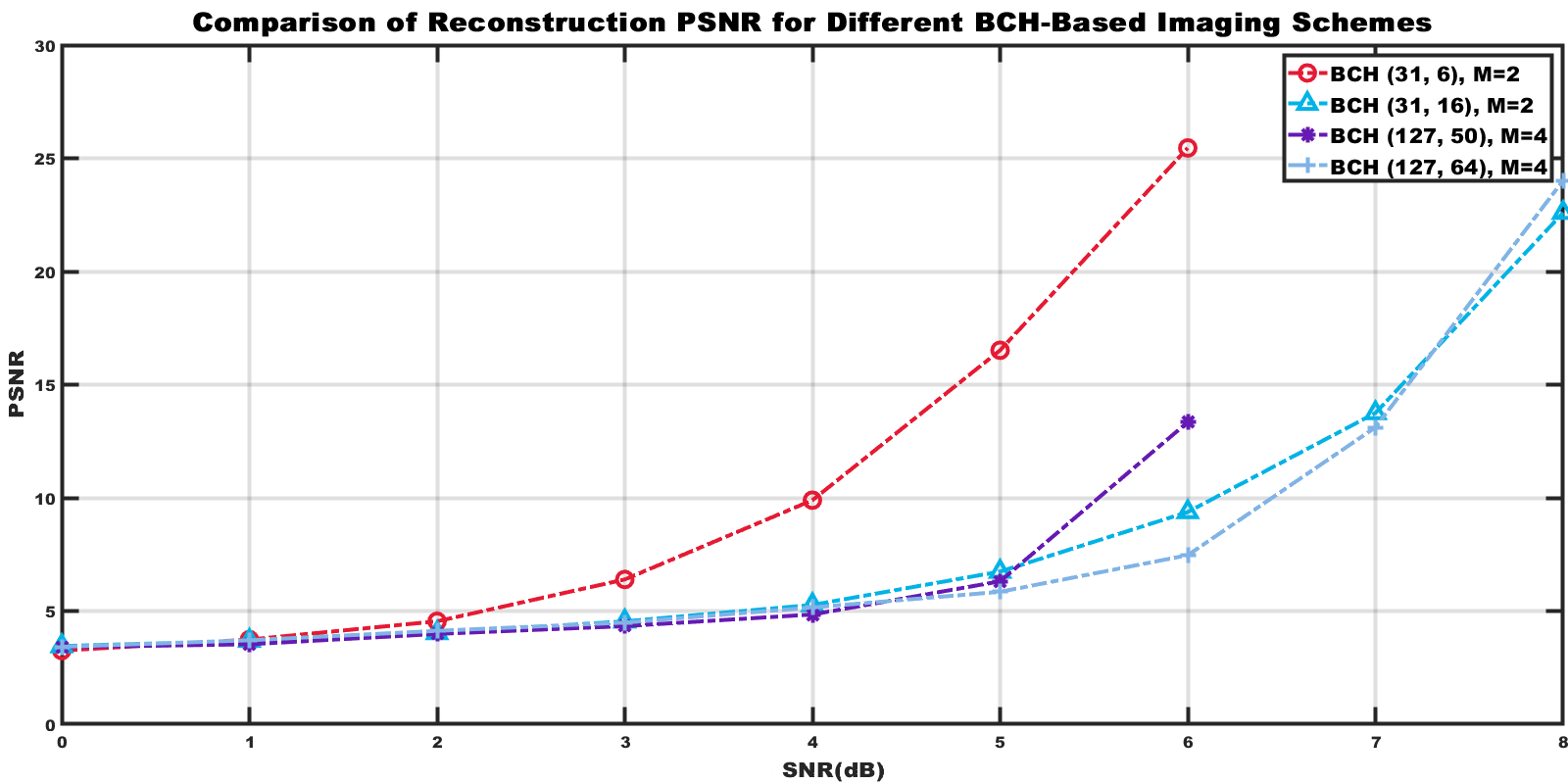}} \\
    \caption{(a). Bit (pixel) error-rate performance of different BCH-based imaging schemes, indicated by four color-coded curves. (b). Reconstruction PSNR, quantifying the reconstruction quality for the four BCH-based imaging schemes.} 
  
    \label{fig:placeholder} 
\end{figure}
predicted for OSD. Analogous to the bit-error metric used in digital communications, we evaluate reconstruction errors by comparing the reconstructed image with the original target on a pixel-by-pixel basis and reporting the resulting pixel error rate.
 Moreover, the simulated \textbf{BCH (31, 16)} imaging results closely follow the OSD theoretical curve in terms of pixel (bit) error rate, indicating that the underlying channel-coding analysis remains valid in the proposed imaging framework. This agreement further confirms that OSD is well suited for decoding the BCH-structured measurements in BCH-based.

\section{Simulation and Experiment}
\subsection{Simulation Performance of Different BCH-based Imaging Schemes}
In our simulations, we consider only AWGN in the bucket detector, while other noise sources are neglected. A \(32\times 32\) blank target image is adopted to evaluate reconstruction performance. We compare four BCH-coded imaging configurations, namely \textbf{BCH (31, 6)}, \textbf{BCH (31, 16)}, \textbf{BCH (127, 50)} and \textbf{BCH (127, 64)}. It is worth noting that, \(M\)-reprocessing with \(M=2\) for OSD is typically sufficient to yield near-ML performance for short block lengths (\(N\le 64\)), whereas longer codes generally require a higher reprocessing order (e.g., \(M\ge 3\) for \(N>64\)) to maintain comparable performance \cite{fossorier1997soft}. \textbf{Fig. 5(a)} and \textbf{Fig. 5(b)} show the bit (pixel) error-rate performance of different BCH-based imaging schemes and the corresponding reconstruction peak signal-to-noise ratio (PSNR), respectively. For a fair and reproducible evaluation across a wide SNR range, the reconstruction quality is quantified by the peak signal-to-noise ratio (PSNR), which is computed from the mean squared error (MSE) between the reference image $\mathrm{T}$ and the reconstructed image $\hat{\mathrm{T}}$:
\begin{equation}
    \mathrm{MSE}=\frac{1}{mn}\sum_{i=1}^{m}\sum_{j=1}^{n}\bigl(T(i,j)-\hat{T}(i,j)\bigr)^2
\end{equation}
\begin{equation}
\mathrm{PSNR}=10\log_{10}\left(\frac{\mathrm{MAX}_T^2}{\mathrm{MSE}}\right),
\end{equation}
where \(\mathrm{m}\) and \(\mathrm{n}\) are the image height and width; $\mathrm{MAX}_T$ denotes the maximum possible pixel value (e.g., 255 for 8-bit grayscale or 1 for binary images). A higher PSNR indicates a smaller reconstruction error and thus better grayscale fidelity.

\textbf{BCH (31, 6)} and \textbf{BCH (31, 16)} share the same block length (\(N=31\)) but exhibit different redundancy levels and error-correction capabilities. Owing to its lower rate (i.e., higher redundancy), \textbf{BCH (31, 6)} provides stronger protection than other three BCH code against random errors and therefore attains higher reconstruction performance under the same noise conditions. A similar tradeoff is observed for the longer codes. \textbf{BCH (127, 50)} and \textbf{BCH (127, 64)} show comparable performance at low-to-moderate SNR (e.g., up to \(6\,\mathrm{dB}\)), whereas the more redundant \textbf{BCH (127, 50)} achieves reliable reconstruction at a lower SNR and improves more rapidly as SNR increases. Increasing the OSD reprocessing order \(M\) can further enhance reconstruction performance for long length BCH code; however, the decoding complexity grows rapidly with \(M\). Although various complexity-reduction techniques and enhanced OSD variants have been reported in the literature, their investigation is beyond the scope of this work. Moreover, \textbf{BCH (31, 16)} and \textbf{BCH (127, 64)} have comparable coding efficiencies, with rates \(R=16/31\) and \(R=64/127\), respectively. Despite their similar coding rates, \textbf{BCH (31, 16)} achieves a slightly lower bit (pixel) error rate and a higher PSNR of reconstruction PSNR than \textbf{BCH (127, 64)}. In the proposed BCH-based imaging scheme, the sampling rate is determined by the code rate \(R\) and can be expressed as \(\hat{R}=\frac{1}{R}\times 100\%\). Moreover, the illumination-pattern size associated with \textbf{BCH (127, 64)} is \(8\times 8\) pixels, which is substantially larger than the \(4\times 4\) pixels' patterns used with \textbf{BCH (31, 16)}. \textbf{Fig. 6} uses a binary "USYD” target to compare the performance of different BCH-based imaging schemes. All schemes can reconstruct the target at low SNR, and the schemes employing BCH codes with stronger error-correction capability generally achieve better reconstruction performance.

It is worth noting that, in BCH-based imaging, increased redundancy alone does not necessarily yield improved performance for longer codes. The benefit of error-control coding depends not only on the nominal error-correction capability, but also on how effectively the redundancy can be exploited by the adopted decoding strategy under finite-measurement constraints \cite{richardson2001efficient}. In addition, excessively small illumination patterns can be impractical. For example, \textbf{BCH (31, 6)} yields a \(32\times32\) illumination-pattern size; as a result, for a \(32\times32\) DMD, each illumination activates at most six micro-mirrors (“sparkling” pixels), which provides limited spatial diversity and makes the configuration more susceptible to model mismatch and experimental perturbations, and therefore less suitable for practical deployment.
\begin{figure*}
    \centering
    \includegraphics[width=1\linewidth]{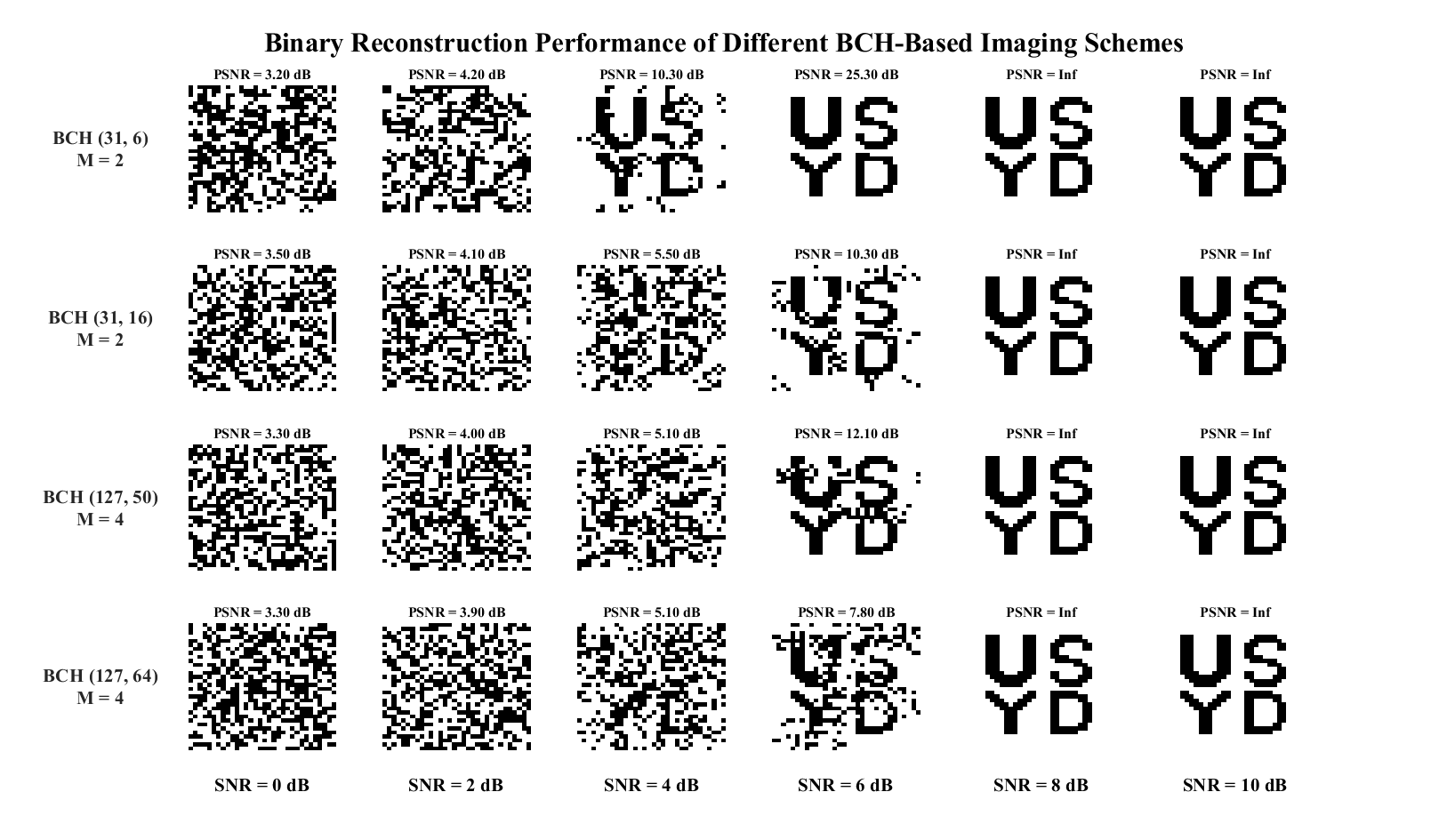}
    \caption{Binary reconstruction performance of different BCH-based imaging schemes under SNR = 0 - 10 dB. PSNR is annotated above each reconstruction (PSNR = Inf indicates an exact reconstruction).}
    \label{fig:placeholder}
\end{figure*}
\subsection{Simulation Performance Comparison Between BCH-Based Imaging Schemes and Second-Order Correlation GI for A Grayscale Target}
To compare the BCH-based schemes with the GI schemes under the same sampling rate, we consider \textbf{BCH (31, 16), \(M =2\)} and \textbf{BCH (127, 64), \(M =4\)}, both operated at \(\hat{R}=200\%\). The conventional second-order correlation GI employed four deterministic illumination patterns: Hadamard, Walsh, CCHDM and Random. Hadamard, Walsh, and CCHDM are widely used illumination families in computational GI. Hadamard patterns form a binary orthogonal basis well suited to DMDs, and generalized sparse Hadamard designs can further improve efficiency and quality under limited measurements~\cite{Han:25}. Walsh patterns share the same orthogonality but enable fast reconstruction via FWHT, achieving fast and high-quality GI~\cite{Wang2016FWHT}. CCHDM is an engineered/re-ordered Hadamard-family set that enhances convergence at low sampling ratios and significantly decreasing the acquisition time \cite{Lopez-Garcia:22}. \textbf{Fig. 7} presents a grayscale-target reconstruction comparison between the proposed BCH-based imaging schemes and conventional second-order correlation GI under several representative illumination pattern families over a wide SNR range. 
As the SNR increases from 0 to 40~dB, all methods exhibit a gradual transition from noise-dominated outputs to visually recognizable structures.

\textbf{Fig.~7} indicates that BCH-based imaging is more robust in the low-to-moderate SNR regime. The BCH schemes recover recognizable structure at lower SNR than correlation GI, suggesting that error-control coding with OSD decoding effectively suppresses measurement noise when GI has not yet converged. Moreover, \textbf{BCH (31, 16)} consistently outperforms \textbf{BCH (127, 64)} across the entire SNR sweep under the current setting. For the GI baselines, deterministic patterns (Hadamard, Walsh, and CCHDM) exhibit similar behavior: the target contour becomes visible at mid-to-high SNR and the reconstructions become stable at high SNR. By contrast, random patterns remain dominated by granular noise and achieve only limited PSNR improvement, consistent with the faster convergence of near-orthogonal deterministic bases under a fixed measurement budget.

Finally, PSNR should be interpreted with the reconstruction domain in mind. GI produces grayscale estimates, whereas BCH-based imaging intrinsically reconstructs \emph{binary} outputs after \(\mathbf{GF^{(2)}}\) mapping and OSD decoding; thus PSNR evaluation requires binarizing the reference (or forming a pseudo-grayscale image by accumulating binary reconstructions). This domain mismatch discards intensity information and can bias PSNR downward, so the lower BCH-based PSNR in \textbf{Fig.~7} primarily reflects the binary reconstruction nature rather than inferior noise suppression.

\begin{figure*}
    \centering
    \includegraphics[width=1\linewidth]{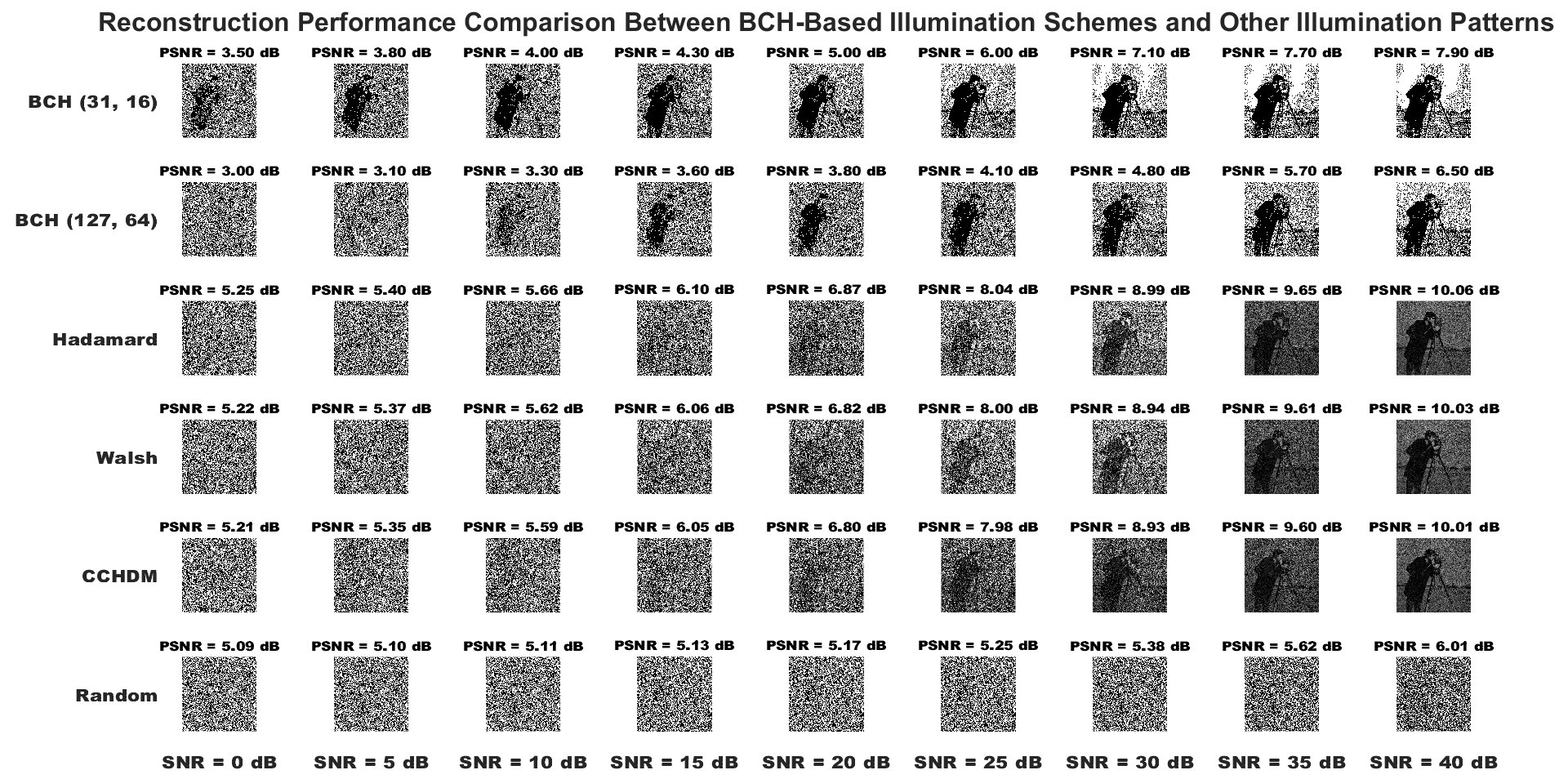}
    \caption{Grayscale reconstruction comparison between BCH-based imaging and second-order correlation GI under different illumination patterns (Hadamard, Walsh, CCHDM, and random) across SNR = 0 - 40 dB.}
    \label{fig:placeholder}
\end{figure*}

\subsection{Experiment Performance Comparison Between BCH-Based Imaging and Second-Order Correlation GI}
    A \(32 \times32\) target is used to test BCH-based imaging system. System configuration and data acquisition is depicted in the system diagram as shown in \textbf{Fig.1(a)}. The imaging setup operates under a centralized synchronous control framework. The optical path begins with a fiber-coupled laser source, followed by optical amplification and collimation. The core component of the system is the DMD, which spatially modulates the beam intensity to project pseudo-random binary patterns onto the target via a projection lens. To ensure high-fidelity reconstruction, a synchronous control module coordinates the timing of the entire system. It simultaneously triggers the modulation of the DMD and the integration of the BD via the acquisition module. The detector collects the total light intensity interacting with the target, compressing the spatial information into a 1D voltage signal. This signal is digitized by the acquisition module and transmitted to a workstation. The final image is computationally recovered by correlating the known DMD patterns with the measured light intensities using BCH decoding algorithm.

We consider two BCH codes, \textbf{BCH (31, 16)} and \textbf{BCH (127, 64)}, which have the same coding rate. \textbf{Fig.~8(a)} illustrates a representative reconstruction example. For a reprocessing order of \(M=2\), \textbf{BCH (31, 16)} yields better reconstruction quality than \textbf{BCH (127, 64)}. When the reprocessing order increases to \(M=4\), the performance of the \textbf{BCH (31, 16)}-based scheme changes only marginally, whereas the \textbf{BCH (127, 64)}-based scheme exhibits a noticeable improvement. These results verify that OSD reprocessing effectively enhances BCH-based imaging performance and enables reliable reconstruction of the target. 
\textbf{Fig.~8(b)} compares the proposed BCH-based imaging scheme with  second-order correlation GI under Hadamard illumination for grayscale reconstruction. Hadamard patterns are known to provide improved reconstruction quality and stronger noise robustness than many alternative illumination patterns \cite{don2019ghost}. To be noted, since the BCH-based imaging scheme reconstructs only binary images, we combine 16 binary reconstructions to form a grayscale image for a fair comparison with GI, which directly reconstructs in grayscale.

Under Hadamard illumination, the sampling rate of GI is set to \(100\%\) and \(200\%\) (relative to the BCH-based sampling budget). Increasing the sampling rate from \(100\%\) to \(200\%\) yields only a marginal performance improvement, rather than a sharp gain. Therefore, we use the \(100\%\) sampling Hadamard-illumination results as the baseline for comparison with the BCH-based imaging scheme. Compared with second-order correlation GI, BCH-based imaging benefits from error-control decoding that mitigates the impact of additive white Gaussian noise (AWGN),which is commonly caused by signal-independent detector noise arising from electronic amplifier noise and thermal noise. 

Meanwhile, the PSNR, which quantitatively measures reconstruction fidelity, is significantly higher for the BCH-based scheme than for conventional GI. This improvement arises because error-control decoding can recover most pixel values exactly, while the remaining errors induced by AWGN tend to be sparse and isolated after decoding. In contrast, second-order correlation GI relies on statistical correlation estimation; consequently, residual noise persists and accumulate gradually as the number of samples increases. Since these errors are sparse and isolated, simple digital post-processing (e.g., median filtering) can further suppress the artifacts and recover the target more completely. This compatibility with standard digital signal processing is another practical advantage of the BCH-based imaging framework. Finally, it is important to note that the remaining reconstruction errors are primarily concentrated near the image center, which can be attributed to nonuniform illumination across the field of view. Improving the illumination uniformity (e.g., by employing a more evenly distributed light source) is therefore expected to further enhance the reconstruction quality.
\begin{figure}
    \centering
\subfloat[(a) ]
    {\includegraphics[width=1\linewidth]{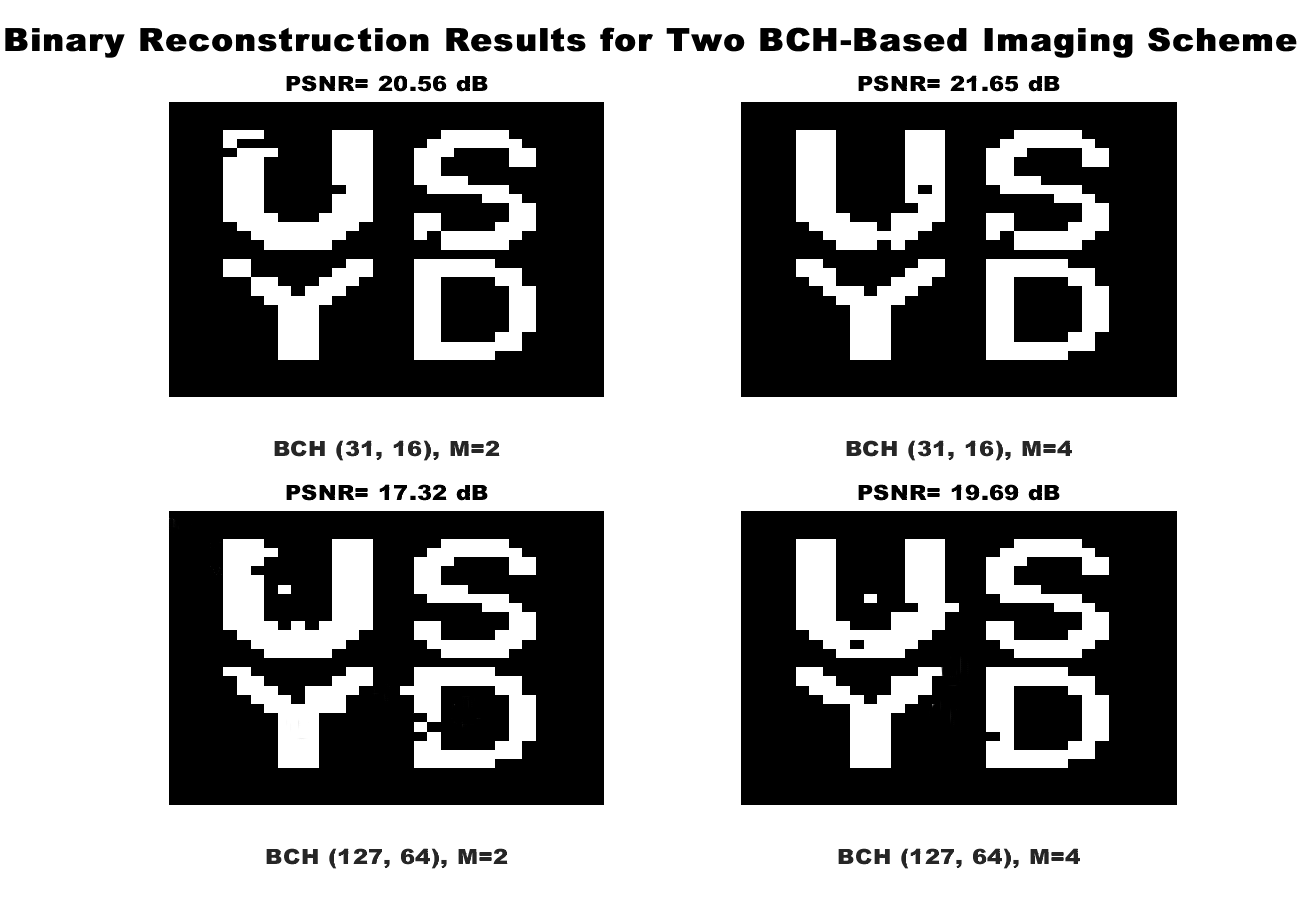}}\\
\subfloat[(b) ]
    {\includegraphics[width=1\linewidth]{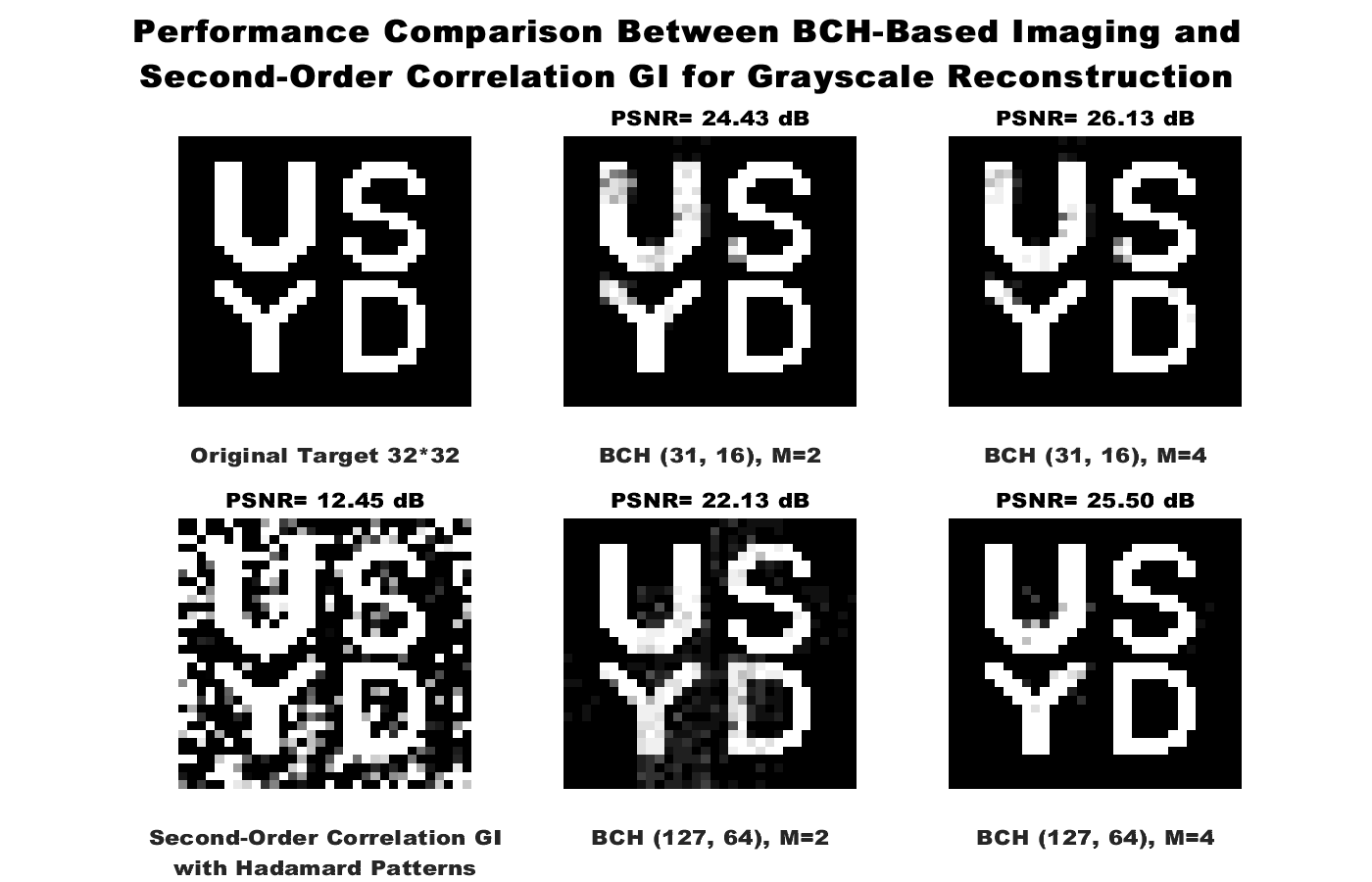}}\\
    \caption{(a). Binary reconstruction results for the \textbf{BCH (31, 16)}-based  and \textbf{BCH (127, 64)}-based imaging schemes with OSD reprocessing orders \(M=2\) and \(M=4\). (b). Performance comparison among \textbf{BCH (31, 16)}-based imaging, \textbf{BCH (127, 64)}-based imaging, and second-order correlation GI for grayscale reconstruction.} 
    \label{fig:placeholder}
\end{figure}
\section{Conclusion}
The application of BCH-based ECC to imaging systems significantly improves their robustness against noise. Through comprehensive theoretical analysis, simulation and experiment, this study demonstrates that BCH codes can be employed to improve image quality improve image quality under AWGN conditions. The results highlight that longer-length BCH codes, such as \textbf{BCH (127, 64)}, strike an effective balance between error-correction capability and coding efficiency for imaging. While the computational complexity increases sharply as the reprocessing order \(M\) grows, exploring more computationally efficient OSD decoding strategies for imaging system is a promising avenue for future research. Future work can focus on optimizing BCH codes for larger image sizes, exploring adaptive coding strategies, and integrating BCH-based techniques with other advanced error control methods to further enhance imaging performance. In addition, comprehensive studies are required to evaluate the impact of two-dimensional ISI, transmission attenuation, and realistic noise models under practical environmental conditions.

\ifCLASSOPTIONcaptionsoff
  
\fi


\bibliographystyle{IEEEtran}
\bibliography{ambcbib}

\end{document}